\documentstyle[fullpage]{article}

\def\del{\partial}

\def\be{\begin{equation}}
\def\ee{\end{equation}}
\def\M{{\widetilde M}}
\def\eq#1{(\ref{#1})}

\begin{document}

\begin{center}
{\large\bf $W_\infty$ Algebra and Geometric Formulation of
QCD$_2$} \\

\medskip

{\large Spenta R. Wadia \\
Tata Institute of Fundamental
Research \\ Homi Bhabha Road, Bombay 400 005, INDIA} \\
\smallskip
{\em e-mail: wadia@theory.tifr.res.in}
\end{center}

Gauge theories and string theory have a long standing symboitic
relationship [1]. In this talk we summarize an application of
some developments in 2-dim. string theory to 2-dim. QCD [2].
These developments are related to $W_\infty$ algebra and its presence
in QCD$_2$ unravels the structure of its non-linear gauge invariant
phase space.  In this framework we will derive 't Hooft's equation
[3] in a geometric setting.

The model is a $U(N)$ gauge theory in 2 space-time dimensions, with
fermions in the fundamental representation of $U(N)$.  Choose the light
cone gauge $A_+(x^-, x^+) = 0$ and consider the gauge invariant
operators, at fixed $x^+{\rm (light-cone~time)}$, involving fermions of
a definite chirality
\be
\label{1}
M(x^-,y^-;x^+) = {1 \over N} \sum^N_{a,b=1} \psi^a_-(x^-,x^+) \:
\left(e^{i\int^{x^-}_x}A_-(z,x^+)^{dz}\right)_{ab}
\psi^{+b}(y^-,x^+)\: \ldots
\ee

In a Hamiltonian formulation at time $x^+ = 0$, we can choose the
gauge $A_-(x^-,0) = 0$.  Then \eq{1} becomes the bilocal operator
$M(x,y) = {1 \over N} \sum^N_{a=1} \psi^a_-(x)\: \psi^{+a}_-(y)$.
Henceforth we will understand $x$ and $y$ as light-cone coordinates
$x^-$ and $y^-$.  Our first observation is that $M(x,y)$ satisfies the
$W_\infty$ algebra
\be
\label{2}
\left[M(x,y),\; M(x',y')\right] =
\delta (x-y')\; M(x',y) - \delta(x' - y) \; M(x,y').
\ee
These are the `Poisson brackets' in the gauge invariant
{\em phase space}.  The second point is that the phase space is
non-linear in the zero charge sector
\be
\label{3}
\int^{+\infty}_{-\infty} M(x,z)\:M(z,y)\:dz  = M(x,y),
\int^{+\infty}_{-\infty} M(x,x)\: dx  =  c
\ee
The constant $c$ is related to the baryon number.  And the third point
is that the Hamiltonian can be expressed entirely in terms of $M(x,y)$:
\begin{eqnarray}
\label{4}
H &=& N \int dxdy \left[{g^2 \over 4}\: M(x,y)\:\tilde M(x,y) -
{im^2 \over 4}\: S(x,y)\: M(x,y)\right] \nonumber \\
\tilde M(x,y) &=& |x-y|\: M(x,y) \;\; {\rm and} \;\; S(x,y) = {\rm
sgn}(x-y)
\end{eqnarray}

The action and the path integral can be constructed using the method
of co-adjoint orbits [4] or the method of $W_\infty$ coherent
states [5]
\be
\label{5}
S = N \Big[2i \int_\Sigma dsdx^+ Tr\left(M\left[\partial_+M, \partial_s
M\right]\right)
 - \int_{-\infty}^{+\infty} dx^+ Tr\left({1 \over 4}\,
im^2 SM + {1\over 4}\, g^2 M \tilde M \right)\Big]
\ee
In \eq{5} $M$ has been treated as an operator with matrix elements
$M(x,y)$ and $\Sigma$ is the half-plane: $(s,x^+) \in (-\infty, 0)
\otimes (-\infty, +\infty)$, with boundary conditions, $M(x,y,x^+,
s=0) = M(x,y,x^+)$ and $M(x,y,x^+, s=-\infty) = 0$.  The phase flow is
described the equation of motion
\be
\label{6}
i\del_+ M  =  (im^2/8)[M,S]+ (g^2/4) \left[M , \M \right]
\ee

A systematic perturbation theory in $1\over N$ can be constructed
using \eq{3},\eq{5} and \eq{6}. In the zero baryon sector, the
classical solution of \eq{3}, \eq{6}, in the momentum basis is
\begin{equation}
\label{7}
M_0 (k,k') = \theta(k) \delta(k - k')
\end{equation}
This solution is related to the filling of the fermi sea upto the
Lorentz invariant (light cone) fermi level $k_F = 0$.  It specifies a
co-adjoint representation of $W_\infty$.  One can say it is the
`master field' of QCD$_2$.

Perturbation theory around $M_0$ is constructed, in analogy with pion
perturbation theory, by the parameterization:
\be
\label{8}
M = e^{iW/\sqrt{N}} M_0\; e^{-iW/\sqrt{N}}
\ee
where $e^{iW/\sqrt{N}}$ is a $W_\infty$ group element,
and $W = \sum_{k,k'} W_{kk'} t_{kk'}$.  The matrices $t_{kk'}$ are the
generators of $W_\infty$ (analogous of the Pauli matrices for
$SU(2)$).  This parameterization preserves \eq{3}.  Now the $W_\infty$
algebra has 2 wedge sub-algebras.  $W_{+\infty} = \left\{t_{kk'}, k >
0, k'>0\right\}$ and $W_{-\infty} = \left\{t_{kk'}, k<0,k'<0\right\}$.
These leave $M_0(k,k')$ invariant and hence the phase space is infact
the coset $W_\infty/W_{+\infty} \otimes W_{-\infty}$, with
co-ordinates $W^{+-} (k,k') = W(k,k'),\: k>0,\: k'<0$ and
$W^{-+}(k,k') = W(k,k')\: k<0,\: k'>0$.  These co-ordinates provide a
non-linear realization of the $W_\infty$ algebra.  Explicit formulas
are obtained by substituting \eq{8} in \eq{2}.  The general algebra of
$W^{+-}$ and $W^{-+}$ is highly non-linear, however to leading order
in $1 \over N$ it reduces to the Heisenberg algebra
\be
\label{9}
{}~[W^{+-}(k,k'), W^{-+}(l,l')] = (1/2)\delta(k-l')\delta(k'-l) +
o(N^{-1/2})
{}~[W^{+-},W^{+-}]=[W^{-+},W^{-+}] = 0 + o(N^{-1/2})
\ee
Substituting \eq{8} into \eq{5} and expanding generates the
perturbation theory $S = NS_0 + S_1 + {1\over N}\: S_2 \dots$.  The
term $S_1$ gives the propagator of the fluctuations $W^{-+}, W^{+-}$
and $S_2,\: S_3$ etc. gives the interactions.

We explicitly present the equation of motion that follows from $S_1$:
\be
\begin{array}{l}
\label{10}
i\del_+ W^{-+}(k, k'; x^+) =
 (m^2/ 4)(1/k + 1/k') W^{-+} (k,k';x^+) \\
\hbox{~~~} -(g^2/ 4\pi)\int_k^{-k'} dp
(1/p^2) \big[W^{-+}(k-p, k'+p;x^+) -
W^{-+}(k,k';x^+) \big] + o(N^{-1/2})
\end{array}
\ee
In \eq{10}, $k>0,\: k'>0$ and we also have the boundary conditions
$W^{-+}(k,k'=0) = 0$ (fermi satisfies) and $W^{-+}(k,k'=\infty)=0$
(finite energy).

Now introducing the variables $x = {k \over r_-},\: r_-=k+k',\: y =
{p_+k' \over r_-}$ and the fourier transform $W^{-+}(k,k';x^+) = \int
{dr^+ \over 2\pi} \phi(x;r_-,r_+) e^{ir_+x^+}$ \eq{10} implies the 't
Hooft equation [3]
\begin{equation}
\label{11}
4r_- r_+ \phi(x) =  m^2( {1\over x} + {1\over 1-x} )\phi(x)
-{g^2\over \pi} \int_0^1 {dy \over (y - x)^2}
\big( \phi(y) - \phi(x))
\end{equation}
with boundary conditions $\phi(0) = \phi(1) = 0$.  This leads to the
well known (stringy) meson spectrum: $\phi_n(x) \sim \sin n\phi x,\:
r_+r_- \sim n$ for large $n$.

\medskip
\noindent {\large Baryons:}
\smallskip

Witten had pointed out that baryons are solitons of the large $N$
theory [6]. In the present framework we have to solve the
eqn. of motion \eq{8} with $M^2 = M$ and $Tr(1-M) = B$.  Their amplitude
is proportional to $e^{-N}$, which is typical of stringy
non-perturbative behaviour [7].  We have to hold the
regularized baryon number to be non-zero.  These non-trivial classical
solutions will correspond to different self-consistent Hartree-Fock
potentials which arise out of populating quasi-particle wave function
above the Dirac sea.  The important point here is that these classical
solutions are given by a function $M_{c\ell} (k^-,k^{-'},t)$ of 2
variables: $k^-\pm k^{-'}$.  If we call the conjugate variables $Y$
and $X$, then the fourier transform of $M_{c\ell}$ is $M_{c\ell}
(Y,X)$.  $X$ represents a centre of mass type co-ordinate.  The
variable $Y$ does not seem to have an analogue in point particle
theories.  It indicates the possibility that the baryon is itself a
{\em stringy} state which appears particle-like only at long wave lengths.
One then expects to introduce 2 collective co-ordinates parameterized
by $(\tau,\sigma): M = m_{c\ell} (Y-y(\sigma,\tau),X-x(\sigma,\tau))$.
One wonders whether these remarks on stringy solitons have any
bearing to those in [8].

\medskip

\noindent {\large Acknowledgement}
\smallskip

I acknowledge A. Dhar and G. Mandal for many discussions on this subject.

\bigskip

\noindent {\bf References}

\begin{enumerate}
\item[{[1]}] {\sl The Large N Expansion in Quantum Field Theory and
Statistical Physics: From Spin Systems to 2-Dimensional
Gravity}, Eds. E. Brezin and S.R. Wadia (World Scientific
1993).
\item[{[2]}] A. Dhar, G. Mandal and S.R. Wadia, Phys. Lett. B 329 (1994) 15.
\item[{[3]}] G. 'tHooft, Nucl. Phys. B72 (1974) 461.
\item[{[4]}] A. Dhar, G. Mandal and S.R. Wadia,
Mod. Phys. Lett. A7 (1992) 3129
\item[{[5]}] A. Dhar, G. Mandal and S.R. Wadia,
Mod. Phys. Lett. A8 (1993) 3557.
\item[{[6]}] E. Witten, Nucl. Phys. B160 (1979) 57.
\item[{[7]}] S.H. Shenker, in Proceedings of the
Cargese Workshop in {\sl Random Surfaces, Quantum Gravity and Strings
1990}, Eds. O. Alvarez, E. Marinari and P. Windey.
\item[{[8]}] J.P. Gauntlet and J.A. Harvey, EFI--94--36, hep-th/9407111.
\end{enumerate}

\end{document}